\documentclass[aps,preprint]{revtex4-1}

\usepackage{graphics}
\usepackage{epsfig}
\usepackage{natbib}
\usepackage{amsmath,amsthm}

\begin{document}

\title{A first-passage-time theory for search and capture of
  chromosomes by microtubules in mitosis}

\author{Manoj Gopalakrishnan}
\email{manoj@physics.iitm.ac.in}
\author{Bindu S. Govindan}
\email{bindu@physics.iitm.ac.in}

\affiliation{Department of Physics, Indian
  Institute of Technology (Madras), Chennai 600036,
  India}


\date{\today}

\begin{abstract}
The mitotic spindle is an important intermediate structure in eukaryotic
cell division, in which each of a pair of duplicated chromosomes is attached through microtubules to centrosomal bodies located close to
the two poles of the dividing cell. Several mechanisms are at work towards the formation of the spindle, 
one of which is the `capture' of chromosome pairs, held together by kinetochores, by randomly searching microtubules. 
Although the entire cell cycle can be up to 24 hours long, the mitotic
phase typically takes only less than an hour. How does the cell keep the duration of mitosis within this limit? 
Previous theoretical studies have suggested that
the chromosome search and capture is optimized by tuning the microtubule dynamic parameters to minimize
the search time. In this paper, we examine this conjecture. We compute the mean search time for a single target by microtubules from 
a single nucleating site, using a systematic and rigorous theoretical 
approach, for arbitrary kinetic parameters. The result is extended to multiple targets and nucleating
sites by physical arguments. Estimates of mitotic time scales are then obtained for different cells using experimental
data. In yeast and mammalian cells, the observed changes in
microtubule kinetics between interphase and mitosis are beneficial in
reducing the search time. In {\it Xenopus} extracts,  by contrast, the
opposite effect is observed, in agreement with the current understanding that large cells use additional mechanisms
to regulate the duration of the mitotic phase.
\end{abstract}

\pacs{05.40.-a, 87.17Aa, 87.16Ln}

\maketitle

{\it Key words}: microtubule, chromosome capture, metaphase spindle, Green's functions

\section{Introduction}

Microtubules are one class of polymeric filaments in the eukaryotic
cell, whose sub-unit is a hetero-dimer of alpha- and
beta-tubulin. Microtubules therefore possess structural polarity, and
the ends are differentiated as plus and minus ends. A hall-mark of
microtubules is their unique mechanism of assembly and dis-assembly: a
polymerizing microtubule can abruptly start shrinking by losing
sub-units and {\it vice-versa}, a process referred to as dynamic
instability(reviewed in \citep{DESAI}).  The stochastic switching process between
growth and shrinkage is referred to as catastrophe and the reverse
process is called rescue. Between a rescue and a catastrophe, a
microtubule grows in length by polymerizing and between a catastrophe
and rescue, it shrinks. In vivo, a third state called pause is also
observed where the length remains static. Microtubules usually
nucleate from organizing centers called centrosomes, but may also be
found free in the cytoplasm.  

Microtubules play a central role in eukaryotic cell division. An
important milestone in the cell division
cycle is the formation of the metaphase spindle, where all the
duplicated chromosome pairs, held together by kinetochores are aligned
along the cell `equator’ (the ``metaphase plate'') in such a way that
each chromosome of a pair
is facing one of the poles of the cell, and attached to one or more
microtubules emanating from a centrosome located near that pole. The spindle starts forming 
when microtubules nucleating from each centrosome randomly searches the surrounding
space for chromosomes by alternately growing and shrinking (the random 
search-and-capture model, and are stabilized upon contact with a kinetochore\citep{MITCHISON, EXP}. 
Investigations over the last decade or so have revealed that the
chromosomes do not always remain passive in this process; rather, the
kinetochores nucleate and stabilize microtubules in their vicinity, a
process facilitated by RanGTP, which then
connect to the astral microtubules emanating from the chromosomes,
assisted by motor proteins such as dynein (see \citep{OCONNELL} for a
recent review). In the present paper, we, however, restrict ourselves
to the situation where chromosomes are passive, and microtubules
perform the search-and-capture.

We now briefly review the previous papers that addressed this problem. 
A theoretical and numerical study of the random search-and-capture model was
done first by Holy and Leibler\citep{HOLY}. In this paper, the conditions for
optimization of search process was investigated for a spherical cell
of radius $R=50 \mu$m, with a single stationary target at various distances $d<R$ from the
centre. The number of searching polymers was fixed at 250, and was
assumed to remain constant with time (effectively infinite nucleation rate). 
At the cell boundary, the filaments would stop growing and wait until a
catastrophe occurs. By a combination of intuitive arguments and
explicit simulations, it was postulated that (i) the global minimum of the
search time occurs when rescue is absent and (ii) the optimized mean time of
search increases with $d$, and is less than 10 minutes for $d<
10\mu$m, but much higher for larger $d$. In a more recent paper, Wollman et al\citep{WOLLMAN} carried out a more
detailed study of the problem, and also investigated the time to capture
multiple chromosomes. An optimal catastrophe frequency was first estimated by
minimizing a weighted average time of search for multiple chromosomes
at variable separations from the nucleating centre. 
Numerical simulations of the problem, using this optimal frequency
showed that the search typically took hours to complete when the number of targets was
large (eg. 46 in humans). However, when a biochemically induced 
bias in search (a microtubule stabilizing RanGTP gradient around chromosomes) was introduced,  the search was completed 
over physiologically reasonable times. 

In the earlier studies, it has generally been assumed, on 
the basis of probabilistic arguments, that the rescue frequency should
be optimally zero. However,  small, but non-zero rescue frequency is
typically observed in mitotic cells, and the existing theoretical
results cannot be used to analyze this case. Also, the earlier studies have generally ignored the 
finite cell size which limits long searches, and is likely to be crucial at least in small cells.   
The number of searching microtubules was generally assumed constant, 
while this is a fluctuating quantity, controlled by the nucleation rate at the centrosomes. 

The primary motivation behind the present paper is to present a
rigorous theoretical method for calculating the search time for 
arbitrary rescue frequency, nucleation rate and cell size/radius. The
formalism presented here is based on a set of Greens functions 
and first passage densities for microtubule dynamics, related through
a set of convolution equations. As such, this approach permits us to
derive an implicit expression for the probability distribution of the 
search time for a single chromosome/target. Existing theoretical
results follow from our more general expressions in the appropriate
limits.We believe that this formalism will be useful in obtaining a
great deal of insight into microtubule dynamics in {\it in vivo}
situations, and may well find applications in other related problems.  

In the following sections, we discuss the problem and the model, develop the formalism
to address the problem, and analyze our results, first in  theoretically interesting limits. 
The results are then discussed in the context of available experimental observations. We
then conclude with a summary of our findings and mention a few
directions in which this study may be extended. Two appendices
supplement the mathematical part of the paper.

\section{Mathematical formalism}

\subsection{Model details}

It is convenient to imagine that during pro-metaphase, prior to division, the shape of
the cell is close to an ellipsoid. Microtubules nucleate from two
centrosomes, which, for simplicity, may be assumed to be located at
the two focal points of the ellipsoid, and the duplicated chromosomes,
held together through kinetochores (henceforth called simply
`targets') are assumed to be scattered around the equatorial plane.  
In the rest of this paper, we only consider capture of the target by
microtubules emanating from one of the centrosomes, which is a
precursor event to the later `bi-oriented' configuration, where
microtubules from both centrosomes will bind to a target and engage in
`tug-of-war' which ultimately separates the individual chromosomes in 
the pair.

It is generally estimated that there are hundreds of nucleating sites in a centrosome. 
From a vacant site, a microtubules nucleate at rate $\nu$  in a random direction 
and grows by polymerization as long as it is in the growth phase, while the same
microtubule shrinks in length by depolymerization in the shrinking
phase. Catastrophe and rescue frequencies are
denoted by $\nu_{c}$ and $\nu_r$ respectively, and are assumed to be the same
everywhere inside the cell, as are the growth and shrinkage
velocities, denoted $v_{g}$ and $v_{s}$. In this paper, as in the
earlier papers which addressed this problem, we will treat
all these different dynamical quantities are independent
parameters (see the discussion in the last section, however). 
In the process, the microtubule scans the surrounding space for
chromosomes, and is stabilized when the growing end encounters a
kinetochore. 

A microtubule from a certain nucleation site on the centrosome can
nucleate in many possible directions; however, given the finite size
of the centrosome, the orientation is likely to be constrained by the
geometry of the centrosome. In an extreme case, one {\bf may} imagine that
a microtubule will always grow only along the local normal to the
surface, but this case is pathological when the target is fixed in
space, since no microtubule might ever grow in the right direction to
find it. It is therefore, more realistic to imagine that microtubules
from each nucleation site in the centrosome will grow within a certain
solid angle $\Delta\Omega$, which defines a {\it search cone} for the
corresponding nucleation site. In this case, if the fixed target falls
inside the cone, and has a cross-sectional area $a$, it subtends a
solid angle $a/d^2$ at a point on the centrosome, and therefore a microtubule
originating at that particular site has a probability $p=ad^{-2}\Delta\Omega^{-1}$ 
for nucleating in the {\it right} direction, within the search
cone of the site.

\begin{figure}
\includegraphics[width=0.8\columnwidth]{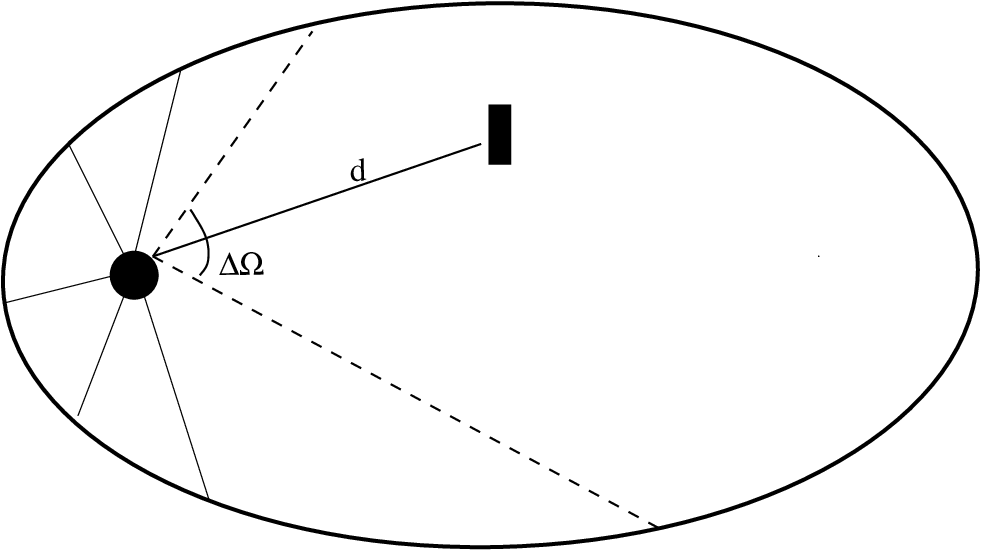}
\caption{\label{fig1} A schematic illustration of the geometry of 
our model is shown here. Microtubules nucleate from nucleating sites
on the centrosome, and search for a stationary target at a distance
$d$. $\Delta\Omega$ is the
solid angle of the `search cone' for a certain nucleating site
depicted in the picture. The search is curtailed by the cell
boundary. The search cones of neighbouring nucleating sites may
overlap (not shown here), which accelerates the process by `parallel' search.}
\label{fig:fig0}
\end{figure}

For a given search process, $\Delta\Omega$ is determined by several factors, the most important being
(a) orientational constraints on nucleation at a given site in the centrosome and (b) steric hindrance 
between microtubules in the cytoplasm. In general, one may see that when $\Delta\Omega$ is large, $p$ is small, 
and consequently, the search by microtubules from any single nucleating site becomes inefficient. However, in this case,
the search cones of different microtubules overlap (each being large)
and more nucleating sites/microtubules will be able to participate in
the search.On the other hand, if $\Delta\Omega$ is small, only a few
microtubules will be effectively searching, 
however the search by each is now more efficient; the two effects therefore compensate each other.
For concreteness, we
choose $\Delta\Omega=\pi/2$ in this paper, i.e., a
microtubule from any nucleating site will be able to search a quarter
of the space around it (This choice is somewhat arbitrary and not directly derived from any experimental 
data; for comparison, Wollman et. al\citep{WOLLMAN} used $\Delta\Omega=2\pi$). An illustration of the model is shown in 
Fig.\ref{fig:fig0}. We take the cross-sectional area of a target to be
$a=0.25\mu$m$^{2}$ throughout this paper (corresponding to a radius of
$\sim 0.28\mu$m), therefore, with the previous estimate of
$\Delta\Omega$, we find $p\simeq 0.31/d^{2}$, 
where $d$ is measured in $\mu m$.For $d=2\mu$m and larger, therefore $p\ll 1$. 

A search by microtubules emanating from any particular site is terminated in three possible ways: if the nucleation
occurs in the right direction, (i) the
growing tip encounters the target and ends the search, or (ii) the
microtubule depolymerizes completely before it encounters the target, and
finally (iii) if the nucleation happens in the wrong direction, the
microtubule will ultimately depolymerize and disappear after a futile
search, with or without encountering the cell boundary. For mathematical simplicity, we assume 
that within a search cone, the boundary is at the same 
distance $R$ from the centre,  although this is strictly true only when $\Delta\Omega$ is sufficiently small. The cut-off distance 
$R$ therefore serves as an estimate of the size of the cell. 
In the last case, we assume that once the microtubule hits the cell boundary, it
undergoes catastrophe at a rate $\nu^{\prime}_c$, which is generally
higher than the value in the interior
\citep{VOROBJEV,GOVINDAN,MOGILNER}. In particular, it was reported in
\citep{VOROBJEV} that the catastrophe frequency near the boundary is
16-fold higher than that in the interior, for certain cells.  
In cases (ii) and (iii), a new
microtubule will nucleate again from the center in a randomly chosen
direction, at rate $\nu$. 

Search cones of many nucleating sites will overlap, and therefore a target will be searched simultaneously by
many microtubules which will reduce the mean search time. It was shown in Wollman et. al\citep{WOLLMAN} that
the mean time to capture $N$ targets by $M$ microtubules ($M$ nucleating sites in our case) is given by

\begin{equation}
T_{M,N}\simeq T_{1,1}\frac{\sigma_N}{M}
\label{eq0}
\end{equation}

where $\sigma_N=\sum_{k=1}^{N}k^{-1}\sim \log N$ when $N\gg 1$.The above expression holds,
provided the probability distribution of the 
search for a single target by a single microtubule is 
a pure exponential (which is true only for zero rescue frequency, as in \cite{WOLLMAN}, but not 
true in general). Further, Eq.\ref{eq0} is true only if 
all the targets are at the same distance from the
centrosome/microtubule nucleating centre, which is not true in a
realistic situation. When the targets are at variable separations, an estimate of
the mean search time can be still obtained in the form of the
following inequality:

\begin{equation}
T_{M,N}< T_{1,1}^{\max}\frac{\sigma_N}{M}
\label{eq0+}
\end{equation}

where $T_{1,1}^{\max}$ is the search time for the farthest target,
which is an upper limit on $T_{1,1}$.  In this paper, we compute $T_{1,1}$ rigorously, with some simplifying
assumptions, but for arbitrary kinetic parameters. We then use 
Eq.\ref{eq0+} to make estimates for multiple targets and parallel
search, for the sake of comparison with experiments.

\subsection{Capture time distribution}

For the rest of this paper, we replace $T_{1,1}$ simply by $T$. 
Let us denote by $C(T)$ the probability density of the capture time
$T$ for a single stationary target at a certain distance from the
centrosome. The mean capture time follows:

\begin{equation}
\langle T\rangle=\frac{\int_{0}^{\infty}dTC(T)T}{\int_{0}^{\infty}dTC(T)},
\label{eq1}
\end{equation}

where, $\int_{0}^{\infty}dTC(T)$ is the probability that the search
will be eventually successful, which we will, later, show to be unity.

Since the basic process under consideration here is the capture of a
certain target by one (or a set) of dynamically unstable filaments, it
is natural to base our theory on consideration of first passage
probability densities\citep{VAN-KAMPEN, REDNER}. For this purpose, it is convenient to 
define a set of three {\it conditional} first passage probability
densities (CFPD), which will serve as the basic quantities in terms of which the probability distribution $C(T)$ 
can be expressed. These CFPDs are defined below, with the corresponding condition for each given in italics.

\begin{enumerate}
\item $K_1(T)\equiv p\Phi(d,T)$, where $\Phi(d,T)$ is the CFPD for a freshly nucleated
microtubule to reach a distance $d$ for the first time after a time interval $T$, {\it without ever shrinking
back to the origin in between}.  

\item $K_2(T)\equiv (1-p)Q_R(T)$, where $Q_X(T)$ is the CFPD for shrinking to
the origin after a life-time $T$, {\it without ever reaching a length $X$ in between}.  

\item $K_3(T)\equiv (1-p)\Psi(T)$, where $\Psi(T)$ is the CFPD for
return to the origin after a time interval $T$,  {\it after encounter with the boundary (and
consequent catastrophe) at least once (and possibly several times) in
between}. 
\end{enumerate}

A successful search event is, in general, preceded by $n$ unsuccessful search events: let us 
denote by $\Omega_n(T)$ the probability of $n$ unsuccessful
nucleation-search-disappearance events within a time interval $T$, so that $C(T)$ may be written as 

\begin{equation}
C(T)=\sum_{n=0}^{\infty}\int_{0}^{T}\Omega_n(T-T^{\prime})p\Phi(d,T^{\prime})\nu
dT^{\prime}
\label{eq2}
\end{equation}

Our next task is to write an expression for $\Omega_n(T)$. 
Let us now assume that among the $n$ unsuccessful nucleation events,
there are $n_1$ events of type 1 (above), where the microtubule nucleated in the right
direction, but did not reach the chromosome, $n_2$ events of type 2 (above), where it
nucleated in a wrong direction, but shrank back to origin before
encountering the boundary, and $n_3=n-n_1-n_2$ events of type 3 (above), where the microtubule
nucleated in a wrong direction, encountered the boundary, underwent
catastrophe and then shrank to the origin. 

Specifying the total number of events in each class does not completely describe the history of the process, as 
the temporal ordering of the events is still arbitrary. 
The $n_1$ events of type 1 can be distributed in a total of $n$ in $\binom{n}{n_1}$ different ways, and
the $n_2$ events of type 2 can be distributed among the remaining $n-n_1$ in $\binom{n-n_1}{n_2}$ different ways. The remaining $n-n_1-n_2$ events 
naturally belong to type 3.  

$\Omega_n(T)$ may now be expressed as a sum over histories (i.e., a {\it path-integral}) of all these events, ordered temporally in all
possible ways. This is done as follows: Starting at time $T=0$, let the first microtubule nucleation occur at time $T_{1}^{\prime}$, and let this
microtubule live for a time interval $T_1$. Then there is a time gap of $T_2^{\prime}$ until the next nucleation, and the microtubule nucleated then
lasts for a time interval $T_2$ and so on. The time gap $T_1^{\prime}$ occurs with a probability $\exp(-\nu T_1^{\prime})$ and the nucleation at the end of 
it occurs with probability $\nu dT_1^{\prime}$. The probability that a microtubule will last for a time interval $T_1$ before shrinking back to the origin may
be denoted $K(T_1)dT_1$, but $K$ could be $K_1$, $K_2$ or $K_3$ depending on whether this event falls into type 1, 2 or 3. For $\Omega_n(T)$, 
there are a total of $n$ such nucleation-death events within a time interval $T$.  The resulting mathematical expression can be written as a convolution over all
these time-intervals, and has the form

\begin{eqnarray}
\Omega_n(T)=\sum_{n_1=0}^{n}\sum_{n_2=0}^{n-n_1}\sum_{per}\int_{0}^{T}\nu dT_{1}^{\prime}e^{-\nu
  T_1^{\prime}}\int_{0}^{T-T_1^{\prime}}dT_1K(T_1).....\nonumber\\
\int_{0}^{T-T_1^{\prime}-..T_{n-1}}
dT_nK(T_n)e^{-\nu[T-\sum_{k=1}^{n}(T_k+T_k^{\prime})]}
\label{eq3}
\end{eqnarray}

where $per$ stands for all the possible permutations of events, as far
as their temporal order of occurrence is concerned. 
The preceding equation has the form of a $2n$-fold convolution, and it is therefore convenient to use Laplace transforms.
We define ${\tilde\Omega_n}(s)=\int_{0}^{\infty}dTe^{-sT}\Omega_n(T)$ and similarly for other quantities. A 
generalized form of the standard convolution theorem for Laplace
transforms may be applied to Eq.\ref{eq3} (see, eg.,\cite{OLF}), and  the result is

\begin{widetext}
\begin{equation}
{\tilde
  \Omega_n}(s)=\frac{1}{(s+\nu)}\bigg(\frac{\nu}{s+\nu}\bigg)^n\sum_{n_1=0}^{n}\sum_{n_2=0}^{n-n_1}\binom{n}{n_1}\binom{n-n_1}{n_2}{\tilde K}_1(s)^{n_1}{\tilde K}_2(s)^{n_2}
  {\tilde K_3}(s)^{n-n_1-n_2}
\label{eq4}
\end{equation}
\end{widetext}

Note that, in the passage from Eq.\ref{eq3} to Eq.\ref{eq4}, allowance has been made for the fact that the random variable $K$ takes the value $K_1$ $n_1$ times,
$K_2$ $n_2$ times and $K_3$ $n-n_1-n_2$ times. The previous equation is clearly a binomial series, and can be summed immediately. From Eq.\ref{eq2}, we find ${\tilde
  C}(s)=\nu p\tilde{\Phi}(d,s)\sum_{n=0}^{\infty}{\tilde
  \Omega}_n(s)$. After substituting the binomial sum from
Eq.\ref{eq4}, and replacing $K_1,K_2,K_3$ by their original notations, we arrive at the following expression:

\begin{equation}
{\tilde C}(s)=\frac{\nu p{\tilde \Phi}(d,s)}{\bigg[s+\nu\bigg(1-p{\tilde
      Q}_d(s)-(1-p)[{\tilde Q}_R(s)+{\tilde \Psi}(s)]\bigg)\bigg]}
\label{eq5}
\end{equation}

As a special case, if MT nucleation occurs very fast and therefore not rate-limiting, we may take the limit $\nu\to\infty$ in the above equation, whence the following 
limiting form is reached:

\begin{equation}
\lim_{\nu\to\infty}{\tilde C}(s)=\frac{p{\tilde \Phi}(d,s)}{1-\bigg[p{\tilde
      Q}_d(s)+(1-p)\bigg({\tilde Q}_R(s)+{\tilde \Psi}(s)\bigg)\bigg]}
\label{eq6}
\end{equation}

Eq.\ref{eq5} is the central result of this paper. Using this
expression, the mean search time for a single target, and its variance
may be expressed as 

\begin{equation}
\langle T\rangle=-\frac{\partial_s{\tilde C}(s)|_{s=0}}{{\tilde C}(0)}~~~;~~~\langle
T^2\rangle=\frac{1}{{\tilde C}(0)}\frac{\partial^2{\tilde C}(s)}{\partial s^2}|_{s=0},
\label{eq7}
\end{equation}

where ${\tilde C}(0)=\int dTC(T)$. It is, however clear that ${\tilde
  \Phi}(d,0)+{\tilde Q}(d,0)=1$, since a microtubule growing in the
right direction will either have to hit the target, or shrink back
without touching the target. Similarly, for the wrong directions, we
have the relation ${\tilde Q}(R,0)+{\tilde \Psi}(0)=1$ for similar
reasons. Substitution of these normalization relations into
Eq.\ref{eq5} shows that ${\tilde C}(0)=1$ for all parameters, i.e.,
the search is always eventually successful. 

The CFPDs introduced above are now calculated from the Green's
functions for MT kinetics, derived explicitly in the next section.

\subsection{Green's functions}

The stochastic state of a MT at a given point in time $t$ is characterized
by two variables, its length $l$ and its state of polymerization
versus depolymerization, which we denote by an index $i$,which takes
one of the two values,1 or 0
respectively for growing and shrinking states. In this case, therefore, we need to compute four Green's
functions, or propagators, $G_{ij}(x,t;x_0,0)$, for $i,j=0,1$; by
definition, $G_{ij}(x,t;x_0,0)dx$ gives the probability that a given
MT will have length $l$ between $x$ and $x+dx$, and will be in state
$i$ at time $t$, provided that it had a length $x_0$ and was at state
$j$ at an earlier time $t=0$. 

Calculating the above Green's functions for a physically realistic
situation would also require specification of appropriate boundary
conditions at the origin (nucleating site) and this has been done
earlier\citep{BICOUT}. However, we deem this unnecessary for our
purpose, since we are only interested in
using these Green's functions to compute the CFPDs introduced
above. Therefore, for the rest of this paper, we will allow the
`length' $x$ to be a continuously varying variable between positive and
negative values, with no boundary condition imposed on the dynamics at
$x=0$. The boundary conditions are used in the definition of the CFPDs later.

The Dogterom-Leibler\citep{DOGTEROM} rate equations for MT kinetics takes the form

\begin{eqnarray}
\partial_tG_{1j}=-v_g\partial_xG_{1j}+\nu_rG_{0j}-\nu_cG_{1j}\nonumber\\
\partial_tG_{0j}=v_s\partial_xG_{0j}+\nu_cG_{1j}-\nu_rG_{0j}
\label{eq8}
\end{eqnarray}

The equations may be solved together using combined Laplace-Fourier
transforms, defined as ${\tilde
  G}_{ij}(k,s;x_0)=\int_{-\infty}^{\infty}e^{-ikx}dx\int_{0}^{\infty}dte^{-st}G_{ij}(x,t;x_0,0)$. The
solution is

\begin{equation}
{\tilde G}_{ij}(k,s)=\frac{e^{-ikx_0}[\nu_r-\delta_{ij}(ikv_s-s)]}{v_sv_g[k^2-ikA(s)+B(s)]}
\label{eq9}
\end{equation}

where 
\begin{eqnarray}
A(s)=[v_s\nu_c-v_g\nu_r+s(v_s-v_g)]/v_sv_g\nonumber\\
B(s)=[s(s+\nu_r+\nu_c)]/v_sv_g.
\label{eq10}
\end{eqnarray}

For connection with the {\bf CFPD}s introduced earlier, it is convenient to
define the Green's function in such a way as that they have dimensions
of inverse time, and not inverse length. This is done by defining

\begin{equation}
F_{1j}=v_gG_{1j}~~;~~F_{0j}=v_sG_{0j}
\label{eq11}
\end{equation}

It will be convenient for later calculations
to carry out the inversion $k\to x$ explicitly:

\begin{widetext}
\begin{eqnarray}
{\tilde
  F}_{1j}(x,s;x_0)=\frac{\nu_r+s\delta_{1j}}{v_s[\alpha_s+\beta_s]}\bigg[e^{-\alpha_s(x-x_0)}\Theta(x-x_0)+e^{\beta_s(x-x_0)}\Theta(x_0-x)\bigg]+\nonumber\\
\frac{\delta_{1j}}{(\alpha_s+\beta_s)}\bigg[\alpha_se^{-\alpha_s(x-x_0)}\Theta(x-x_0)-\beta_se^{\beta_s(x-x_0)}\Theta(x_0-x)\bigg]\nonumber\\
{\tilde F}_{0j}(x,s;x_0)=\frac{\nu_c+s\delta_{0j}}{v_g[\alpha_s+\beta_s]}\bigg[e^{-\alpha_s(x-x_0)}\Theta(x-x_0)+e^{\beta_s(x-x_0)}\Theta(x_0-x)\bigg]-\nonumber\\
\frac{\delta_{0j}}{(\alpha_s+\beta_s)}\bigg[\alpha_se^{-\alpha_s(x-x_0)}\Theta(x-x_0)-\beta_se^{\beta_s(x-x_0)}\Theta(x_0-x)\bigg]\nonumber\\
\label{eq12}
\end{eqnarray}
\end{widetext}

where

\begin{eqnarray}
\alpha_s=\frac{A(s)}{2}+\sqrt{B(s)+A^2(s)/4}\nonumber\\
\beta_s=-\frac{A(s)}{2}+\sqrt{B(s)+A^2(s)/4}
\label{eq13}
\end{eqnarray}

and $\Theta(x)$ is the usual step-function: $\Theta(x)=1$ for $x\geq
0$ and 0 otherwise.

\subsection{Calculation of $\Phi(d,T)$ and $Q_X(T)$}

The Green's functions calculated in the last section may now be used to
compute the CFPDs which we used before. For this purpose, it is
convenient to define first a set of {\it unconditional} first passage
densities (denoted FPD) as follows: let $C_{ij}(x,t;x_0,0)$ denote the
probability, per unit time, for a MT in state $j$ and with length
$x_0$ at time $t=0$, to reach a length $x$ for the first time at time
$t$, and in state $i$. 

For $l>0$, $C_{11}(l,t;0,0)$ is given by the implicit equation

\begin{eqnarray}
F_{11}(l,t;0,0)=C_{11}(l,t;0,0)+\nonumber\\
\lim_{\epsilon\to 0+}\int_{0}^{t}dt^{\prime}C_{11}(l,t^{\prime};0,0)F_{11}(l-\epsilon,t;l,t^{\prime})
\label{eq14}
\end{eqnarray}

In the above equation (and the following equations), the
$\epsilon$-factors take into account the following restriction on its
dynamics: starting from a growing state at $t=0$, with a length $l$,
it can return to the same length $l$ at a later time, in a growing
state, only from below (which decides which of the $\Theta$-functions
appearing in Eq.\ref{eq12} is non-zero). 
Similar restrictions apply to the equations below.

Similarly, $C_{01}(0,T;d,0)$ and $C_{10}(d,T;0,0)$ are given by the equations

\begin{eqnarray}
F_{01}(0,T;d,0)=C_{01}(0,T;d,0)+\nonumber\\
\lim_{\epsilon\to
  0+}\int_{0}^{T}dT^{\prime}C_{01}(0,T^{\prime};d,0)F_{00}(\epsilon,T;0,T^{\prime})\\
\label{eq15}
F_{10}(d,T;0,0)=C_{10}(d,T;0,0)+\nonumber\\
\lim_{\epsilon\to
  0+}\int_{0}^{T}dT^{\prime}C_{10}(d,T^{\prime};0,0)F_{11}(d-\epsilon,T;d,T^{\prime})
\label{eq15+}
\end{eqnarray}

Using these two FPDs, we are now in a position to write down the
following relations between the CFPDs introduced earlier:

\begin{equation}
C_{11}(d,t;0,0)=\Phi(d,t)+
\int_{0}^{t}dt^{\prime}Q_{d}(0,t^{\prime})C_{10}(d,t;0,t^{\prime})
\label{eq16}
\end{equation}

\begin{equation}
C_{01}(0,T;0,0)=Q_d(T)+\int_{0}^{T}dT^{\prime}\Phi(d,T^{\prime})C_{01}(0,T;d,T^{\prime})
\label{eq17}
\end{equation}

Eq.\ref{eq14}-\ref{eq17} may now be solved using Laplace
transforms. From Eq.\ref{eq14}, we find that

\begin{equation}
{\tilde C}_{11}(d,s;0)=\lim_{\epsilon\to 0+}\frac{{\tilde
    F}_{11}(d,s;0)}{1+{\tilde F}_{11}(d-\epsilon,s;d)}=e^{-\alpha_sd}
\label{eq18}
\end{equation}

Similarly,

\begin{equation}
{\tilde
  C}_{10}(d,s;0)=\frac{\nu_re^{-\alpha_sd}}{\nu_r+s+\alpha_sv_s}~~;~~{\tilde C}_{01}(0,s;d)=\frac{\nu_ce^{-\beta_sd}}{\nu_c+s+\beta_sv_g}
\label{eq19}
\end{equation}

After solving Eq.\ref{eq16} and Eq.\ref{eq17} together, and using Eq.\ref{eq18},\ref{eq19}, we find the
explicit expressions

\begin{eqnarray}
{\tilde
  \Phi}(d,s)=\frac{D(s)e^{-\alpha_sd}}{\nu_r\nu_c[1-e^{-(\alpha_s+\beta_s)d}]+D(s)}\nonumber\\
{\tilde Q}_d(s)=\frac{\nu_c}{\nu_c+s+\beta
  v_g}\left[1-e^{-\beta_sd}{\tilde \Phi}(d,s)\right]
\label{eq20}
\end{eqnarray}

where 

\begin{equation}
D(s)=(s+\alpha_sv_s)(s+\beta_sv_g)+\nu_r(s+\beta_sv_g)+\nu_c(s+\alpha_sv_s).
\label{eq21}
\end{equation}

\subsection{Calculation of $\Psi(T)$: Catastrophes at the cell boundary}

We assume that when a MT hits the cell boundary by growing, it
undergoes catastrophe there at a rate $\nu_c^{\prime}$. We now compute
$\Psi(T)$, which is the CFPD of return to origin (i.e., complete
depolymerization) of a MT after a lifetime $T$, and an encounter with
the boundary at least once.

Clearly, along the line of our previous arguments, $\Psi(T)$ may be given
by the expression

\begin{eqnarray}
\Psi(T)=\int_{0}^{T}
dT_1\Phi(R,T_1)\times\nonumber\\
\int_{0}^{T-T_1}\nu_c^{\prime}dT_2e^{-\nu^{\prime}_cT_2}\chi(R,T-T_1-T_2),
\label{eq22}
\end{eqnarray}

where $\chi(R,T)$ gives the FPD of complete depolymerization of a MT,
starting at the boundary, at length $R$ in shrinking state, with
possibly multiple visits back to the boundary in between. This
quantity may now be expressed implicitly through the equation

\begin{eqnarray}
\chi(R,T)=\Phi^*(R,T)+\int_{0}^{T}dT_1Q_R^*(T_1)\times\nonumber\\
\int_{0}^{T-T_1}\nu^{\prime}_cdT_2e^{-\nu^{\prime}_cT_2}\chi(R,T-T_1-T_2)
\label{eq23}
\end{eqnarray}

where $Q_R^*(T)$ is a `mirror' image, or {\it dynamic inverse} of the quantity $Q_R(T)$ introduced earlier,
and represents the CPFD of a return to boundary over a time interval
$T$, without ever reaching the origin (i.e., shrinking to zero) in
between. Similarly, $\Phi^{*}(R,T)$ is the `inverse' of $\Phi(R,T)$,
and gives the CFPD of complete depolymerization of a MT starting at
the boundary, without ever returning to the boundary in between.

Eq.\ref{eq22} and Eq.\ref{eq23} may now be solved together using Laplace
transforms, and we find

\begin{equation}
{\tilde\Psi}(s)=\frac{\nu^{\prime}_c{\tilde
    \Phi}(R,s){\tilde \Phi^*}(R,s)}{s+\nu^{\prime}_c\bigg(1-{\tilde Q}_R^*(s)\bigg)}
\label{eq24}
\end{equation}

The inverse quantities ${\tilde \Phi^*}(R,s)$ and ${\tilde Q}_R^*(s)$ may be obtained from
${\tilde \Phi}(R,s)$ and $Q_R(s)$ respectively by the transformations
$v_s\leftrightarrow v_g$ and $\nu_r\leftrightarrow\nu_c$. From Eq.\ref{eq13},
this has the effect of replacing $\alpha_s$ by $\beta_s$ and
{\it vice-versa}, while $D(s)$ defined in Eq.\ref{eq21} remains invariant
under these transformations. Therefore, using Eq.\ref{eq20}, we arrive at the following
expressions for these `inverse' quantities:

\begin{eqnarray}
{\tilde
  Q_R^*}(s)=\frac{\nu_r}{\nu_r+s+\alpha_sv_s}\bigg[1-e^{-\alpha_sR}{\tilde
    \Phi^*}(R,s)\bigg]\\
{\tilde
  \Phi^*}(R,s)=\frac{D(s)e^{-\beta_sR}}{\nu_r\nu_c[1-e^{-(\alpha_s+\beta_s)R}]+D(s)}
\label{eq25}
\end{eqnarray}

Eq.\ref{eq25} completes the list of quantities that we need to compute
the mean search time. We now start from Eq.\ref{eq7}, and after a few elementary calculations and
rearrangement of terms, it turns out that the complete expression for
the mean search time may be written out as follows(see Appendix A for details):

\begin{equation}
\langle T\rangle=N_s[pt_d+(1-p)t_R+t_{\nu}]=T_d+\frac{1-p}{p}T_R+T_{\nu}
\label{eq26}
\end{equation}

where $N_s=[p{\tilde \Phi}_d(0)]^{-1}$ gives the mean number of unsuccessful search events before each successful
one, and $pt_d+(1-p)t_R+t_{\nu}$ gives the weighted mean lifetime per event. $t_d=-{\tilde \Phi}_d^{\prime}(0)-
{\tilde Q}^{\prime}(d,0)$ gives the mean time of a search in the right direction, the first term corresponding to the single successful 
search event and the second giving the mean of all unsuccessful events. $t_R=-{\tilde Q}_{R}^{\prime}(0)-{\tilde \psi}^{\prime}(0)$ is the mean 
lifetime of an unsuccessful search event in the wrong direction, the first term corresponding to events not reaching the boundary while 
the second corresponds to events which hit the boundary at least
once. Note that $t_R$ is the mean lifetime of
microtubules. Finally, $t_{\nu}=\nu^{-1}$ is the mean time between the
disappearance of one microtubule and nucleation of a new one at a
site, and is the only term that depends on the nucleation rate $\nu$.

Various quantities such as the mean, standard deviation and higher
moments (if necessary) of the search time, as well as other quantities
such as the mean lifetime of microtubules may now be calculated using
the set of equations presented in this section, and parameters such
as catastrophe and rescue frequencies as well as growth and shortening
velocities taken from experiments. We found it convenient to use
Mathematica (Version 7, Wolfram Research) to carry out the explicit
computations. The results will be discussed in the following section.

\section{Results}

In this section, we will analyze some experimental observations
using the results from our model.Experimental measurements of the microtubule kinetic parameters show
that distinct changes occur as the cell progresses from interphase to
mitosis\citep{RUSAN}, see Table I. Budding Yeast cells show a reduction in both
catastrophe and  rescue frequencies, but the changes are relatively small. 
In mammalian cells, which are typically larger, there is a marked fall in rescue
frequency between interphase and mitosis, and a two-fold increase in
catastrophe frequency. In {\it Xenopus} oocytes (frog egg cells, which
are large and almost 1 mm in radius), the
effects are somewhat different: the rescue frequency, while
small, is almost doubled, but more remarkably, there is a sharp, seven-fold rise in the 
catastrophe frequency.

Given that mitosis occupies only a small fraction of the total time duration of a cell cycle, we 
first seek to determine whether the changes in microtubule kinetics are beneficial to reduce the mean time
of search. For all cases discussed below, we choose $\nu_c^{\prime}=10\nu_c$ to be roughly 
consistent with experiments\citep{VOROBJEV}. However, reducing or increasing $\nu_c^{\prime}$ by an order of magnitude 
does not significantly affect the results.

\subsection{Yeast and mammalian cells show features that are
  consistent with the random search and capture model, but {\it
    Xenopus} oocytes do not}

\begin{figure}[h!]
\includegraphics[width=0.8\columnwidth]{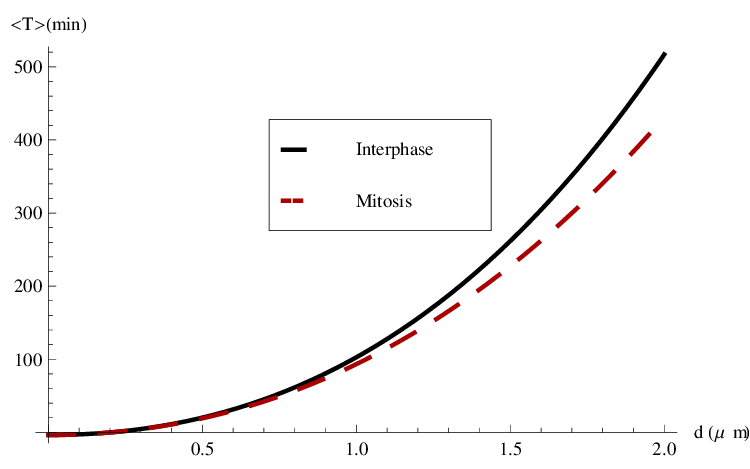}
\caption{The mean time of capture (in minutes) of a
  single chromosome in budding yeast, by microtubules from a single nucleating
  site, for various target distance $d$ is shown here. The thick black
  like corresponds to interphase values of $\nu_r$ and $\nu_c$. We
  assume the nuclear radius to be $R=2\mu m$. The other parameters are 
  $v_d=3\mu$m min$^{-1}$, $v_g=2\mu$m min$^{-1}$, 
$\nu=0.1$min$^{-1}$ and $\nu_c^{\prime}=10\nu_c$.}
\label{fig:fig1}
\end{figure}

Fig.\ref{fig:fig1} shows a comparison for the mean time of search, between interphase and mitosis values in
yeast, for a range of target distance $d$. Since yeast undergoes closed mitosis, the relevant boundary cut-off length scale 
is the nuclear radius, which we take to be $R=2\mu$m.  The mitosis parameters clearly reduces the time scale relative to the
interphase parameters, though understandably, given the small cut-off radius of search, the effect is small.  
When the centrosomal microtubule nucleation frequency is chosen to be $\nu=0.1$min$^{-1}$ per site (see next paragraph)
$\langle T\rangle$ in yeast is found to vary from 100-400 minutes, for $d$ between 1$\mu$m and 
2$\mu$m. Taking $T_{1,1}^{\max}\sim 400$ min, and using $N=32$ in
budding yeast, we see from Eq.\ref{eq0+} that the mean time to
complete search is 24 min, with 50 searching microtubules.

We now turn to the case of mammalian cells. Experiments by Piehl et. al\cite{NUCLEATION} measured a nucleation rate of $\sim$ 80-100 min$^{-1}$, per centrosome 
in kidney epithelial (LLCPK) cells. The total number of nucleation sites is unknown, but from the measured surface area of centrosomes in metaphase ($\sim$ 100 $\mu$m$^{2}$), and
the base area of a single microtubule ($\pi r^2\simeq 4\times 10^{-4}\mu$m$^{2}$, with $2r=25nm$), a total of almost $10^{5}$ nucleation sites are possible, which is clearly 
too large; as a conservative estimate, we may assume a total of 1000 nucleation sites. Then, the nucleation rate per site may be roughly estimated as $\nu\simeq 0.1$min$^{-1}$. 

\begin{figure}[h!]
\includegraphics[width=0.8\columnwidth]{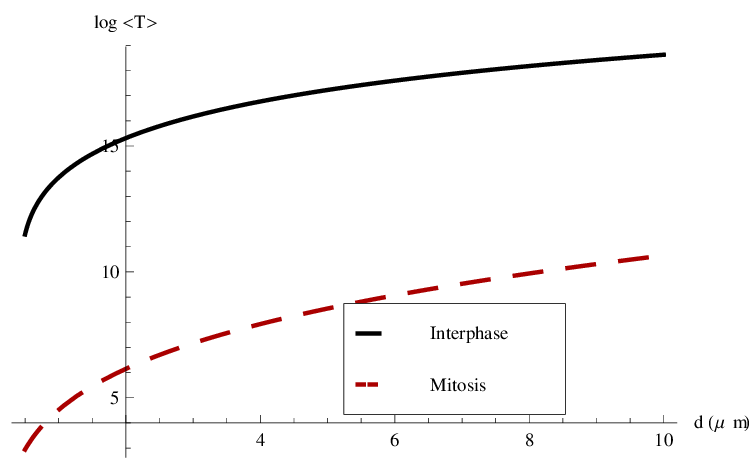}
\caption{Similar to the previous figure, but for mammalian cells. The
  cell radius is taken as $R=20\mu$m. The other parameters are 
  $v_d=14\mu$m min$^{-1}$, $v_g=12\mu$m min$^{-1}$, 
$\nu=0.1$min$^{-1}$ and $\nu_c^{\prime}=10\nu_c$.
}
\label{fig:fig2}
\end{figure}

Let us now perform our analysis on a mammalian cell of radius
$R=20\mu$m, and compare the search time between interphase and
mitosis. 
As seen in Fig.\ref{fig:fig2}, where the logarithm of the time is shown against the distance $d$, the mitosis values significantly reduce the mean search time, almost by 4 orders of magnitude!  For $d=6\mu$m, the time 
computed from mitosis values is $T_{1,1}^{\max}\simeq 2000$ minutes. 
If we conservatively assume that at least $M=100$ kinetochore
microtubules will be actively searching for one target, and 
since $N>10$ typically (46 in humans), using Eq.\ref{eq0+}, we arrive at an estimate of 35-40 minutes for the total
mean search time, which is reasonable. However, this is only the
average time, and may not represent a typical value. We have observed that the standard deviation of $T$ is 
typically of the same order as $\langle T\rangle$; therefore, in an individual experiment, the search could take twice as long.

\begin{figure}[h!]
\includegraphics[width=0.8\columnwidth]{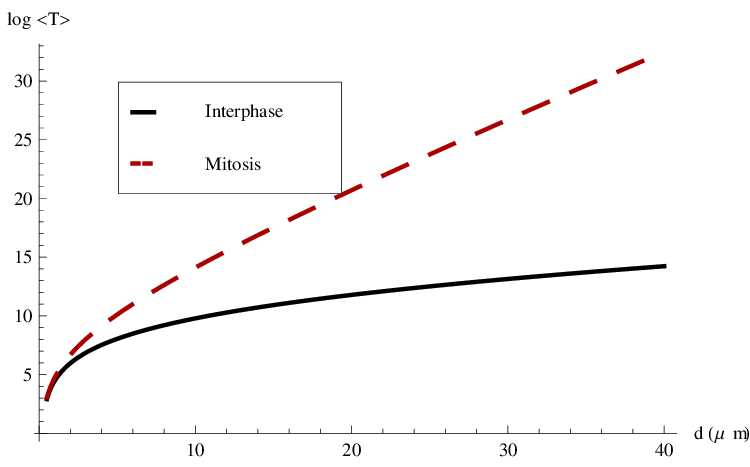}
\caption{Similar to the previous figures, but for {\it Xenopus}
    oocyte cells. The
  cell radius is taken as $R=500\mu$m. The other parameters are 
  $v_d=15\mu$m min$^{-1}$, $v_g=12\mu$m min$^{-1}$, 
$\nu=0.1$min$^{-1}$ and $\nu_c^{\prime}=10\nu_c$. Note that, unlike
  the previous figures, mitosis values appear to increase the search
  time compared to interphase, which suggests that the random search and capture mechanism might 
be inefficient in large cells.
}
\label{fig:fig3}
\end{figure}

In {\it Xenopus} extracts, the situation is very different (see Fig.\ref{fig:fig3}). In this case, surprisingly, it is the interphase parameter values that give the lower mean search time, and mitosis time is typically 1-2 orders of magnitude larger! Even the interphase mean search time is quite large, ranging from 2000-10,000 minutes for $d$ between 5 and 10 $\mu$m, and it would 
need almost 1000 actively searching microtubules to bring the total search time down to acceptable values. The anomalously large value of the catastrophe  frequency in mitosis is puzzling, since targets that are far are more effectively searched when $\nu_c$ is small. . How do we understand
this discrepancy? One reason, as is now becoming increasingly clear,
could be that the random search
and capture mechanism is simply inefficient in such large cells. 
Indeed, it is now well-established that large cells like oocytes require additional mechanisms of 
search like actin contractile ring\cite{ACTIN} and guided search
through RanGTP gradient around chromosomes\cite{OHBA,WOLLMAN}. In the latter
case, microtubules are preferentially stabilized close to chromosomes
hy a gradient of the protein Ran, and the increased catastrophe
frequency might simply serve to enhance the turnover rate, and hence
the dynamicity of microtubules.

\subsection{Mathematical analysis show that $\nu_r=0$ is a minimum condition when $R\gg d$}

In this section, we take a closer look at the theoretical expression
for the mean search time derived earlier, and try to understand some
general features, from the point of view of optimization of the search
process, i.e., minimization of the search time. Without any loss of
generality, we may put $\nu=\infty$ here, as the only effect of a
finite $\nu$ is to add a timescale of $\nu^{-1}$ to the mean time (see Eq.\ref{eq26}).

It is instructive to look at the mean search time separately for parameter regimes 
demarcated by the conditions $A(0)>0$ and $A(0)<0$. Physically, these conditions correspond to
a bounded growth regime for the filaments, with finite mean length and unbounded growth regime with 
mean length linearly increasing with time\citep{DOGTEROM}. After
rather lengthy algebraic calculations, an explicit expression for
$\langle T\rangle$ can be obtained, and the terms therein could be classified into two: (i) those that remain finite,
or diverge as $R\to\infty$ and (ii) those that involve only terms of the form
$e^{-R|A(0)|}$ which disappear for large $R$. In the following, we
ignore terms that fall into (ii), since they are not crucial for our analysis here.

{\bf \it Bounded growth regime (BG); $A(0)>0$}:

In this case, ${\tilde \Phi}(d,0)=D_0e^{-A(0)d}/[D_0+\nu_r\nu_c(1-e^{-A(0)d})]$, so that 
the mean number of trials $N_s$ diverges exponentially with $d$. The
breakup of the total time in the limit $R|A(0)|\to\infty$, is as below:


\begin{eqnarray}
T_d=\frac{\beta^{\prime}+v_g^{-1}}{A(0)}[e^{A(0)d}-1]-\beta^{\prime}d\nonumber\\
T_R\simeq\bigg (1-\frac{\nu_r v_g}{\nu_c v_s}e^{-A(0)d}\bigg)\frac{e^{A(0)d}\nu_c v_s(v_s+v_g)}{(v_s v_g A(0))^2}
\label{eq27}
\end{eqnarray}

where $\beta^{\prime}=(\nu_r+\nu_c)/v_sv_gA(0)$. In this regime, 
$T_d$ and $T_R$ diverge exponentially with $d$, but only linearly with $R$, with the $R$-dependence entirely disappearing
for $R\gg A(0)^{-1}$ (terms now shown in the equation). 
This is physically reasonable, since $A(0)^{-1}$ is indeed the mean length of the filaments in this regime. We also
observe that $T_R$ diverges as $A(0)^{-2}$ and $T_d$ diverges as $A(0)^{-1}$ as $A(0)\to 0$. The intuitive explanation is that, at this
`critical point' in parameter space, the tip of a microtubule performs a pure one-dimensional random walk, whose mean time of return to 
the origin is infinite, as is well-known. 

The $R|A(0)|\gg 1$ limit, with some further simplifications, is treated in more detail in Appendix B. We will now give the corresponding results for the case $A(0)<0$. 


{\bf \it Unbounded growth regime(UBG): $A(0)<0$}:

In this case, ${\tilde \Phi}(d,0)=D_0/(D_0+\nu_r \nu_c[1-e^{-|A(0)|d}]$, which is finite as $d\to\infty$. This is natural, because under conditions of 
unbounded growth, the microtubules are able to reach out to larger distances with fewer attempts compared to the bounded growth regime.
Similar to the previous case, we may now
calculate the breakup of the terms in the limit $R|A(0)|\gg 1$ and $\nu_c^{\prime}\to\infty$. The results are given below:

\begin{eqnarray}
T_d=\frac{|A(0)|}{\nu_r v_g}F(d)-\frac{\nu_c v_s}{\nu_r v_g}\beta^{\prime}d+\frac{\nu_c(v_s-v_g)}{v_g^2v_s |A(0)|^2}[1-e^{-|A(0)|d}]\nonumber\\
T_R\simeq \frac{\nu_r
  v_g}{H(d)}\bigg\{\frac{v_s}{\nu_r}|A(0)|F(R)+F^{*}(R)-
\frac{\nu_c v_s}{\nu_r
  v_g}\beta^{\prime}R+\frac{\nu_c(v_s-v_g)}{v_g^2v_s
  |A(0)|^2}+\nonumber\\
\frac{v_g+v_s}{v_g v_s |A(0)|}\bigg[e^{|A(0)|R}\bigg(1+\frac{\nu_r}{v_s|A(0)|}\bigg)-1\bigg]\bigg\}
\label{eq28}
\end{eqnarray}

where 

\begin{eqnarray}
H(d)=\nu_c v_s-\nu_r v_g e^{-|A(0|d}\nonumber\\
F(R)=-\frac{D^{\prime}}{D_0}+\frac{D^{\prime}}{D_0+\nu_r\nu_c}+\alpha^{\prime}R\nonumber\\
F^*(R)=F(R)+(\beta^{\prime}-\alpha^{\prime})R\nonumber\\
\alpha^{\prime}=\frac{\nu_r+\nu_c}{v_g v_s |A(0)|}~~~;~~~\beta^{\prime}=\frac{v_s^2\nu_c-v_g^2\nu_r}{(v_s v_g)^2 |A(0)|}
\label{eq29}
\end{eqnarray}

In contrast to the previous case, $T_R$ diverges exponentially with $R$, while $T_d$ diverges only linearly with $d$. As $|A(0)|\to 0$, 
the mean time again diverges as $|A(0)|^{-2}$ as before. The exponential divergence with $R$ arises solely from the boundary-interaction term ${\tilde \Psi}^{\prime}(0)$; the
return-to-origin term ${\tilde Q}^{\prime}(0)$ has a finite limit as $R\to\infty$.This result is in quantitative agreement with the corresponding results in \citep{BICOUT} for
the mean time of return to origin of a microtubule in unbounded (as well as bounded) growth phase.

The analogy between the dynamics of the tip of a microtubule and a
one-dimensional biased random walk as developed in \citep{BICOUT} is
helpful in understanding these results. It has been shown that the bias (drift) of this random walk 
is proportional to $v_g\nu_r-v_s\nu_c$, i.e., the bias is negative when (in our notation)
$A(0)>0$ (`bounded growth phase'), positive when
$A(0)<0$ (`unbounded growth phase') and the walk is unbiased (i.e.,
the tip moves diffusively) when $A(0)=0$. In the limit $R\to\infty$, $T_R$ and $T_d$ diverge as $A(0)\to 0$ simply because 
the mean time of return to origin of a one-dimensional random walk is infinite, whereas this time is finite for a biased random walk.

\begin{figure}[h!]
\includegraphics[width=0.7\columnwidth]{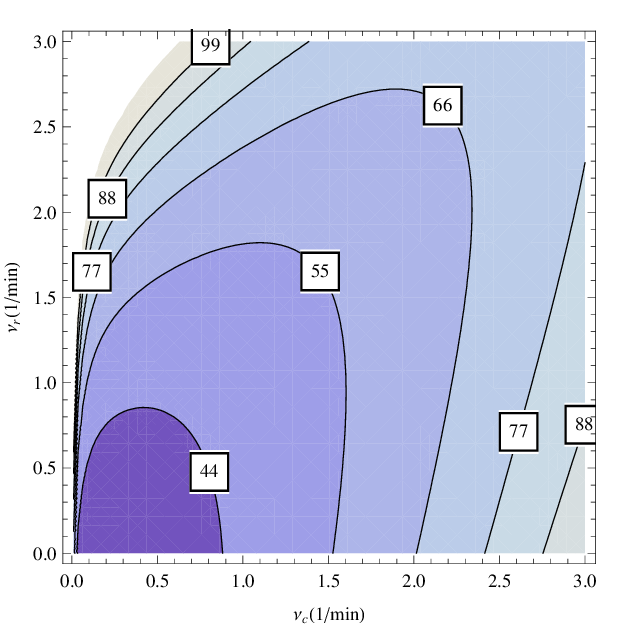}
A.
\includegraphics[width=0.7\columnwidth]{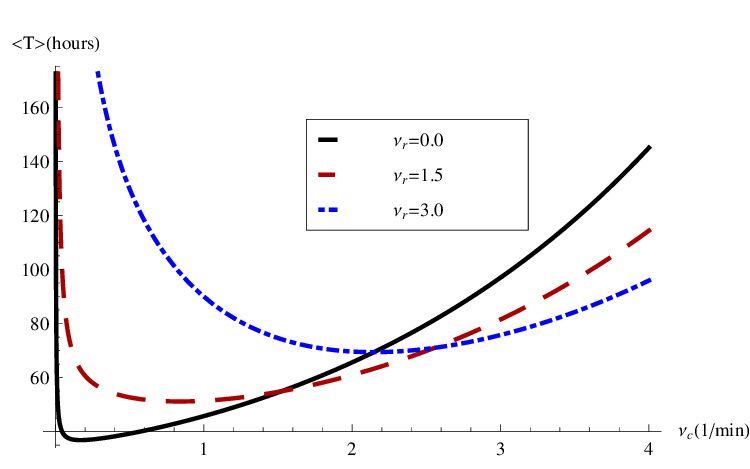}
B.
\caption{A. Contour plot for the mean search time (expressed in hours) of a
  single chromosome at $d=6\mu$m, by a single nucleating site,  in a mammalian cell with radius $R=20\mu$ m. B. 
  Cross-sections of the same plot at three values of $\nu_r$. The parameters are 
  $v_d=14\mu$m min$^{-1}$, $v_g=12\mu$m min$^{-1}$, 
$\nu=0.1$min$^{-1}$ and $\nu_c^{\prime}=10\nu_c$.}
\label{fig:fig4}
\end{figure}

\begin{figure}[h!]
\includegraphics[width=0.8\columnwidth]{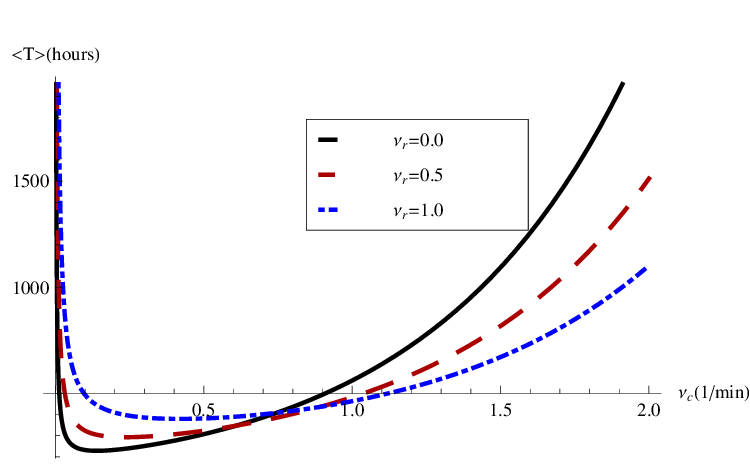}
\caption{The mean time of search for a target close to the boundary, at $d=20\mu$m, in a cell
  of radius $R=20\mu$m for three different $\nu_r$. The
  other parameters are chosen as $v_s=12\mu$m min$^{-1}$, $v_g=10\mu$m
  min$^{-1}$, $\nu=1$min$^{-1}$  and $\nu_c^{\prime}=10\nu_c$. Note
  that slightly larger values of $\nu_r$ produce a more robust minimum
  as a function of $\nu_c$, at the cost of a small increase in time.
}
\label{fig:fig5}
\end{figure}

We may now make a few general observations from these results.  For a single target, as considered here, it is appropriate to assume that $p\ll 1$, in which case, the $T_R$ term dominates over $T_d$ in Eq.\ref{eq26}. 
If $R\gg d$, the exponential divergence of $T_R$ with $R$ in the UBG
regime makes it less favourable compared to the bounded growth regime. For the
latter case, at least when $R\gg A(0)^{-1}$, it is proved in Appendix
B that $T_R$ is a monotonically increasing function of $\nu_r$, i.e.,
it is minimized for $\nu_r=0$. Therefore, we conclude that if the cell boundary is sufficiently far
in comparison with $d$, for a single target, search is optimal if the
microtubules are in a bounded growth phase, at zero rescue frequency. 
This conclusion is in agreement with the previous authors\cite{HOLY,
  WOLLMAN}.Interestingly, these conclusions hold even if $R$ is only
about thrice as large as $d$, as shown in Fig.\ref{fig:fig4}, or even
when $d=R$ (Fig.\ref{fig:fig5}).

In a more realistic situation where a number of chromosomes are
distributed randomly in the cytoplasm at varying distances, 
$p$ itself effectively becomes a dynamic variable, starting at a large value and
progressively decreasing with time as targets are captured one by one. In this
situation, it is likely that search is optimized at a small, but non-zero
rescue frequency. Preliminary results shown in
Fig.\ref{fig:fig5} suggest that when the target is far from the
centrosome, non-zero rescue does not increase the mean search time
significantly, and in addition, could produce a more robust minimum. 

\subsection{Non-zero rescue is likely to be a compromise between
  $\nu_r=0$ and $\nu\to\infty$}

If the search time is minimized at zero rescue frequency, as shown by the previous arguments, why is not the observed rescue frequency in mitosis {\it even smaller}? 
We believe that this could possibly
reflect a compromise between minimizing rescue and maximizing nucleation. Both rescue frequency and nucleation rate depend directly on the concentration of free
tubulin in cytoplasm. Experimental observations by Walker et. al.\cite{WALKER} have shown that rescue frequency is an almost linearly increasing function of free GTP-tubulin
concentration, and nucleation rate is an even more strongly increasing function of concentration. Therefore, the observed rescue frequency in mitosis could probably be understood as the result of  a more general optimization exercise also involving nucleation and catastrophe frequencies, as well as growth velocity, all of which depend on free
GTP-tubulin concentration. 

\subsection{Microtubule turnover time is much smaller in mitosis}

\begin{figure}[h!]
\includegraphics[width=0.7\columnwidth]{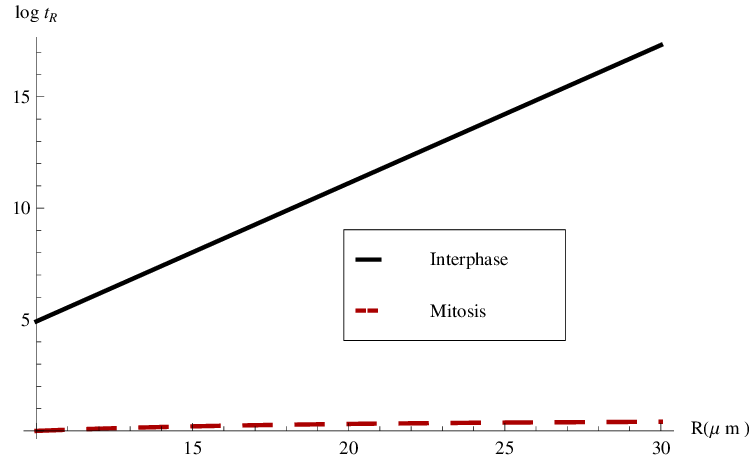}
A.
\includegraphics[width=0.7\columnwidth]{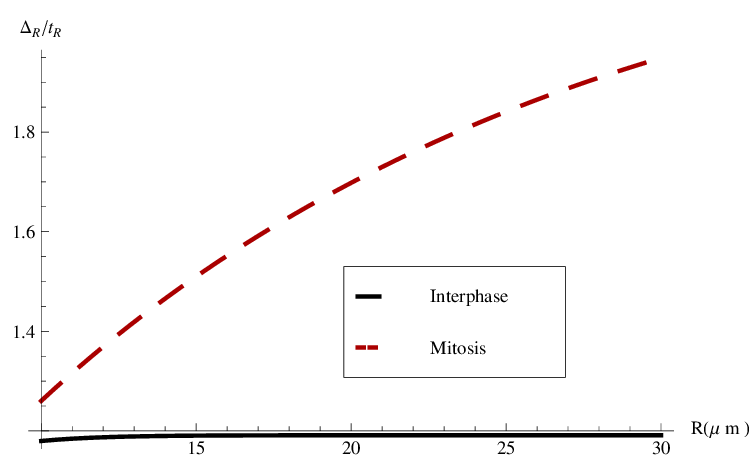}
B.
\caption{A. The mean lifetime $t_R$, defined as the mean lifetime of
  microtubules nucleating in directions away from the target (Eq.\ref{eq26}), is
  plotted as a function of the cell radius $R$ for interphase and
  mitotic parameter values. The lifetime in interphase is larger
  by several orders of magnitude.The parameter values are chosen as in
  Fig.\ref{fig:fig3}. B. The relative fluctuation in the
  lifetime, defined as the ratio of standard deviation $\Delta_R$
  to the mean $t_R$, as a function of $R$, for the same set of parameter
  values. Fluctuations in the mitotic phase are larger than in
  interphase, signaling increased dynamicity of the microtubules.}
\label{fig:fig6}
\end{figure}

In Fig.\ref{fig:fig6}A, we show the mean lifetime (defined
in Eq.\ref{eq26}) of microtubules searching in the wrong directions, as
a function of cell size $R$, in both interphase and mitosis. The
lifetime in mitosis is several orders of magnitude smaller than
interphase, and varies little with cell size (a direct consequence of
being in the BG regime discussed earlier). However, mitotic
microtubules are also more dynamic: the standard deviation of the
lifetime as a fraction of the mean, is larger in mitosis compared
to interphase (Fig.\ref{fig:fig6}B). Experimental observations in
mammalian cells have shown that microtubule turnover in mitosis is 18-fold higher than in
interphase\cite{SAXTON}.

\section{Conclusions}

In this paper, we studied the capture of a target by dynamically
unstable microtubules using a novel and mathematically rigorous 
first passage time-based formalism. Compared to earlier studies, the
principal new features in our model are (a) estimation of the mean time of capture at
non-zero rescue frequency (b) introduction of the cell size as a
parameter in the theory and (c) explicit comparison with experimental
observations in different mitotic cells. Although the model was formulated for the
purpose of understanding chromosome capture in mitosis, the formalism
itself is very general. In particular, the technique 
could be directly applied to the study of cortical capture of microtubules(see, eg.\cite{LEE}) and other similar problems.

Several {\it in vivo} experiments have shown distinct and significant
changes in microtubules dynamics in different cells, as the cell
proceeds from interphase to mitosis. We sought to determine whether
these changes are beneficial to the search and capture of
chromosomes. Our analysis shows that in yeast and mammalian cells, the
mean search time for a single target is reduced in mitosis compared to
interphase. In {\it Xenopus} oocytes, by contrast, the experimental observations could not be
reconciled with the observed changes in microtubule dynamics between
interphase and mitosis, suggesting that the basic strategy of search may be
strongly modified by additional mechanisms.

Although this was not our main interest, we also tried to determine theoretically the conditions for
minimization of the mean search time. We showed rigorously that when
the target is well inside the boundary, the time is minimized at zero
rescue and an optimal catastrophe frequency, in agreement with
previous authors. However,when the target is close to the boundary, although $\nu_r=0$ is still the absolute minimum, 
it was observed that a small, but non-zero $\nu_r$ produced a
more robust minimum (with respect to change in $\nu_c$), and therefore
could be preferred by cells. 

The present study was only concerned with a single target, while in all realistic situations, multiple chromosome 
pairs have to be captured. Unfortunately, to extend the present
analysis to multiple targets would require more detailed knowledge of
$C(T)$, but given the complexity of the mathematical form for ${\tilde
  C}(s)$, this is not too easy(see, however, \citep{WOLLMAN}, where
this analysis was done for exponentially decaying $C(T)$ at
$\nu_r=0$). We are presently working on extracting information about $C(T)$ from our formalism, and extending our
analysis for multiple targets. In particular, it is not immediately clear whether the optimization
criteria for multiple targets will be the same as for a single target,
especially when multiple targets are at variable separations from the
centrosome.  Further, as discussed earlier, the various dynamic
parameters for microtubule dynamics are not generally independent (eg. nucleation and 
rescue could be related, and detailed GTP cap theories suggest that
growth velocity is related to catastrophe frequency\citep{FLYVBJERG}).
A more general optimization scheme has to take these possibilities
into account, and could produce a non-zero optimal rescue
frequency. We leave these ideas to a future study.

Other possible extensions of this study involve
including (i) chromosome diffusion (ii) side-capture of microtubules by
chromosomes through intermediaries like kinesin-13 motor
proteins, followed by one-dimensional diffusive or directed motion to
the tip\citep{HOWARD} and (iii) microtubule nucleation close to the chromosomes.
The latter is a possible alternative to the end-capture mechanism
studied in this paper and it would be interesting to look at its
effect on the mitotic time-scales and its optimization.

\begin{acknowledgments} 
BG acknowledges financial support under a Young Scientist-Fast Track
Research Fellowship, from the Department of Science and Technology, Government of India. MG
would like to thank Mohan Balasubramanian (NUS, Singapore) for a brief but illuminating
conversation on mitosis in yeast. MG's work was supported in part by a
grant from the Centre for Industrial Consultancy and Sponsored Research, IIT Madras. The authors
are thankful for the hospitality under the Visitors Program at the Max-Planck Institute for
Physics of Complex Systems, Dresden, Germany, where this work was
initiated. 
\end{acknowledgments}

\section{Appendix I}

{\it General expression for the mean search time}:

After a series of calculations, the following general
expression is reached for the mean search time from Eq.\ref{eq5}.

\begin{equation}
\langle T\rangle=T_d+\frac{1-p}{p}T_R+\frac{1}{p}T_\nu
\label{eqA1}
\end{equation}

with

\begin{eqnarray}
T_d=-[{\tilde \Phi}(d,0)]^{-1}\left[{\tilde\Phi}^{\prime}(d,0)+{\tilde
    Q^{\prime}}(d,0)\right]\nonumber\\
T_R=-[{\tilde \Phi}(d,0)]^{-1}\left[{\tilde Q^{\prime}}(R,0)+{\tilde
    {\Psi}^{\prime}}(0)\right]\nonumber\\
T_\nu=\frac{1}{\nu }[{\tilde \Phi}(d,0)]^{-1}
\label{eqA2}
\end{eqnarray}

where ${\tilde \Phi^{\prime}}(d,0)=\partial_s{\tilde
  \Phi}(d,s)|_{s=0}$, ${\tilde Q^{\prime}}(d,0)=\partial_s{\tilde
  Q}(d,s)|_{s=0}$,${\tilde Q^{\prime}}(R,0)=\partial_s{\tilde
  Q}(R,s)|_{s=0}$ and ${\tilde \Psi^{\prime}}(0)=\partial_s{\tilde
  \Psi}(s)|_{s=0}$. Here, $T_d$, $T_R$ and $T_\nu$ represent,
respectively, the mean time spent in 
searching in the right direction, wrong directions and between successive nucleations.We note
that the last term disappears in the (theoretical) $\nu\to\infty$ limit, where the
nucleation happens infinitely fast. Also, for small $p$, $T_R$ and
$T_\nu$ dominate over $T_d$, since, in this limit, it is the
unsuccessful search events that take up most of the time spent on
search. 

The exact analytical forms for these functions are as given below:

\begin{widetext}
\begin{eqnarray}
{\tilde\Phi}^{\prime}(X,0)={\tilde\Phi}(X,0)\left[\frac{D^{\prime}}{D_0}-\frac{D^{\prime}+\nu_r\nu_c(\alpha^{\prime}+\beta^{\prime})Xe^{-\gamma_0X}}{D_0+\nu_r\nu_c(1-e^{-\gamma_0X})}-\alpha^{\prime}X\right]\nonumber\\
{\tilde Q}^{\prime}(X,0)=-{\tilde
  Q}(X,0)\left[\frac{1+\beta^{\prime}v_g}{\nu_c+\beta_0v_g}+e^{-\beta_0X}\frac{[{\tilde
    \Phi}^{\prime}(X,0)-\beta^{\prime}X{\tilde
        \Phi}(X,0)]}{[1-{\tilde\Phi}(X,0)e^{-\beta_0X}]}\right]\nonumber\\
{\tilde \Psi}^{\prime}(0)={\tilde \Psi}(0)\left[\frac{{\tilde
      \Phi}^{\prime}(R,0)}{{\tilde \Phi}(R,0)}+\frac{{\tilde
      \Phi}^{*\prime}(R,0)}{{\tilde
      \Phi}^{*}(R,0)}-\frac{[1-\nu_c^{\prime}{\tilde
      Q}^{*\prime}(R,0)]}{\nu_c^{\prime}[1-{\tilde
        Q}^{*}(R,0)]}\right]
\label{eqA3}
\end{eqnarray} 
\end{widetext}

and 

\begin{eqnarray}
{\tilde \Phi}^{*\prime}(R,0)={\tilde \Phi}^{\prime}(R,0;v_g\rightarrow
v_s,\nu_r\rightarrow \nu_c)\nonumber\\
{\tilde Q}^{*\prime}(R,0)={\tilde Q}^{\prime}(R,0;v_g\rightarrow
v_s,\nu_r\rightarrow \nu_c)
\label{eqA4}
\end{eqnarray}

are the mirror-image quantities defined in text. The time-integrated
probabilities ${\tilde \Phi}(d,0),{\tilde Q}(X,0)(X=d,R)$ and ${\tilde
  \Psi}(0)$ are given by

\begin{eqnarray}
{\tilde
  \Phi}(X,0)=\frac{D_0e^{-\alpha_0X}}{D_0+\nu_r\nu_c[1-e^{-\gamma_0X}]}\nonumber\\
{\tilde Q}(X,0)=\frac{\nu_c}{\nu_c+\beta_0v_g}\left[1-{\tilde
    \Phi}(X,0)e^{-\beta_0X}\right]\nonumber\\
{\tilde \Psi}(0)=\frac{{\tilde \Phi}(R,0){\tilde
  \Phi}^{*}(R,0)}{1-{\tilde Q}^{*}(R,0)}.
\label{eqA5}
\end{eqnarray}

The cofficients appearing in the above expressions are defined as
follows:

\begin{eqnarray}
\alpha_0\equiv
\alpha(0);\alpha^{\prime}=\partial_s\alpha(s)|_{s=0}\nonumber\\
\beta_0\equiv
\beta(0);\beta^{\prime}=\partial_s\beta(s)|_{s=0}\nonumber\\
D_0\equiv D(0);D^{\prime}=\partial_sD(s)|_{s=0}
\label{eqA6}
\end{eqnarray}


and $\theta=\partial_{s}A(s)|_{s=0}=(v_s-v_g)/v_sv_g$, $\theta^{\prime}=\partial_{s}B(s)|_{s=0}=(\nu_r+\nu_c)/v_sv_g$.



\section{Appendix II}

{\it Some simple special cases in BG regime}

If $p\ll 1$, $T_R$ and $T_\nu$ dominate over $T_d$, and, from
Eq.\ref{eq26}  and Eq.\ref{eq27} we have

\begin{eqnarray}
\langle
T\rangle\simeq\frac{e^{A(0)d}}{p}\left(1+\frac{\nu_r(1-e^{-A(0)d})}{v_sA(0)}\right)\left[\frac{v_s+v_g}{v_sv_gA(0)}+\frac{1}{\nu}\right]
\nonumber\\(p\ll
1, R\to\infty)~~~~~~
\label{eqB1}
\end{eqnarray}

It can be shown that this expression is a monotonically increasing
function of $\nu_r$, where $0\leq \nu_r<(v_g/v_s)\nu_c$ in the BG regime. In
order to see this, let us define $x=\nu_cv_s$ and $y=\nu_rv_g$, and
$x-y>0$ in this regime. Then, in terms of $x$ and $y$, the mean search
time may be expressed in the form

\begin{equation}
\langle T\rangle_{x,y}=\frac{[xe^{\delta(x-y)}-y]}{(x-y)}\left[\frac{a_1}{(x-y)}+b_1\right]\geq \frac{a_1}{(x-y)}+b_1
\label{eqB1+}
\end{equation}

where $a_1$, $b_1$ and $\delta$ are positive constants, and the
inequality follows because $e^{\delta(x-y)}\geq 1$. The lower bound in
the above equation continuously increases with $y$ in the
applicable range [0:$x$]. We then conclude that $\langle
T\rangle$ itself is an increasing function of $y$, and hence $\nu_r$. Therefore, $\nu_r=0$ is a
necessary condition for a minimum. In this case, Eq.\ref{eqB1}
reduces to

\begin{equation}
\langle
T\rangle=\bigg[\frac{1}{\nu}+\frac{1}{\nu_c}(1+v_g/v_s)\bigg]\frac{e^{\frac{\nu_cd}{v_g}}}{p}-\frac{d}{v_s}~~~(p\ll
1, R\to\infty,\nu_r=0)
\label{eqB2}
\end{equation}

which is minimized $\nu_c=\nu_c^{\min}$, where

\begin{equation}
\nu_c^{\min}=\frac{2v_g}{d\bigg[1+\sqrt{1+\frac{4v_g}{d\nu(1+v_g/v_s)}}\bigg]}~~~~(p\ll
1,R\to\infty,\nu_r=0)
\label{eqB3}
\end{equation}

Finally, in the limit $\nu\to\infty$, $\nu_c^{\min}=v_g/d$, and the optimized search time is

\begin{equation}
\langle T\rangle^{\min}=\Gamma
d^3-\frac{d}{v_s};~~~~ (p\ll 1,R\to\infty,\nu_r=0,\nu\to\infty)
\label{eqB4}
\end{equation}

where $\Gamma=e\Delta\Omega/[a(v_g^{-1}+v_s^{-1})]$ from Eq.\ref{eq0}.

The expression in Eq.\ref{eqB2} above may be approximately reproduced
by physical arguments as follows\citep{HOLY}. The probability that a microtubule
will nucleate in the right direction, and will not undergo catastrophe
until it
reaches the target is given by $p_s=pe^{-\nu_cd/v_g}$, and it will
take at least $N\sim p_s^{-1}$ unsuccessful attempts before this is
accomplished. Each of these unsuccessful search events lasts a time
$\tau\sim \nu_c^{-1}$, and therefore, the total search time is 

\begin{equation}
T_s\sim N\nu_c^{-1}=\frac{e^{\nu_cd/v_g}}{p\nu_c}.
\label{eqB5}
\end{equation}

Note that Eq.\ref{eqB2} reduces to Eq.\ref{eqB5} in the limit $p\to
0$, $\nu\to\infty$ and $v_g\ll v_s$.

\newpage

\begin{table}
\begin{tabular}{cccc}
\hline
\hline
     & Budding Yeast (I) & mammalian (II) & {\it Xenopus} extracts(III) \\
\hline
$\nu_r$ & (0.42)0.12 min$^{-1}$ & (10.5)2.7 min$^{-1}$ & (0.66)1.2 min$^{-1}$\\
$\nu_c$ & (0.48)0.24 min$^{-1}$ & ((1.56)3.48 min$^{-1}$ & (1.08)7.2 min$^{-1}$\\
$v_g$   & 1.7 $\mu$m min$^{-1}$ & 12.8 $\mu$m min$^{-1}$ & 12.3  $\mu$m min$^{-1}$\\
$v_s$   & 2.7 $\mu$m min$^{-1}$ & 14.1 $\mu$m min$^{-1}$ & 15.3 $\mu$m
min$^{-1}$\\
$R$ & 2$\mu$m & 20$\mu$m & 500$\mu$m\\
\hline
\end{tabular}
\caption{Experimental values of microtubule kinetics in the mitotic
  phase. The values in parantheses are interphase values, prior to
  the cell entering mitosis, for Yeast\citep{TIRNAUER}, mammalian\citep{RUSAN} and
  {\it Xenopus} extracts\citep{BELMONT}. The cell radii given are only
  rough estimates. For 
theoretical and numerical analysis, we used $v_g=2.0$ and $v_s=3.0$ $\mu$m
min$^{-1}$ for I, and $v_g=12.0$ and $v_s=14.0$ $\mu$m
min$^{-1}$ for II and III. The experimental data for different cell
sizes are summarized in \citep{RUSAN}.}
\end{table}

\newpage

\newpage

\newpage

\newpage

\newpage

\newpage

\newpage

\end{document}